# Characterisation of a silicon photomultiplier device for applications in liquid argon based neutrino physics and dark matter searches

P.K. Lightfoot [a*], G.J. Barker [b], K. Mavrokoridis [a], Y.A. Ramachers [b], N.J.C. Spooner [a]

[a] Department of Physics and Astronomy, University of Sheffield, Hicks Building, Hounsfield Road, Sheffield, S3 7RH, UK
[b] Department of Physics, University of Warwick, Coventry, CV4 7AL, UK

**Abstract**

The performance of a silicon photomultiplier has been assessed at low temperature in order to evaluate its suitability as a scintillation readout device in liquid argon particle physics detectors. The gain, measured as $2.1 \times 10^6$ for a constant over-voltage of 4V was measured between 25°C and -196°C and found to be invariant with temperature, the corresponding single photoelectron dark count rate reducing from 1MHz to 40Hz respectively. Following multiple thermal cycles no deterioration in the device performance was observed.

The photon detection efficiency (PDE) was assessed as a function of photon wavelength and temperature. For an over-voltage of 4V, the PDE, found again to be invariant with temperature, was measured as 25% for 460nm photons and 11% for 680nm photons.

Device saturation due to high photon flux rate, observed both at room temperature and -196°C, was again found to be independent of temperature. Although the output signal remained proportional to the input signal so long as the saturation limit was not exceeded, the photoelectron pulse resolution and decay time increased slightly at -196°C.

*Corresponding author:* Dr. Phil Lightfoot, Department of Physics and Astronomy,

University of Sheffield, Hicks Building, Hounsfield Road, Sheffield S3 7RH, UK,

Tel: +44(0)1142224533, Fax: +44(0)1142223555, E-mail:p.k.lightfoot@sheffield.ac.uk

# 1. Introduction

Typically liquid noble gas detectors (XENON [1], ZEPLIN [2], ICARUS [3]) rely on quartz photomultiplier tubes (PMTs) to detect UV scintillation light. The large cost, complexity of readout, and relatively high intrinsic background [4] of PMTs has led us to study other optical readout options for future scintillation detectors. Requirements are high quantum efficiency (~20%), fast rise-time (~3ns), low dark count rate (~50Hz) and high gain (~$10^7$). In such detectors measurement of the smaller primary ionization charge signal is essential both to allow characterization of the interaction and to obtain positional sensitivity. Any PMT replacement must therefore be capable of similar performance.

The silicon photomultiplier (SiPM) [5,6] is a recent evolution of the avalanche photodiode (APD) and the large area avalanche photodiode (LAAPD) [7,8,9,10]. Compact, cheap and of high radiopurity, with low power consumption and supply voltage requirements, and capable of reasonably high gain and reduced noise at low temperatures, SiPM devices [11] are now being considered for low temperature photon readout applications both in medium [12] and high energy physics [13]. SiPMs are particularly appropriate for use in scintillation detection because of their high sensitivity, high quantum efficiency, and insensitivity to magnetic fields up to 4T [6]. Excellent time [14] and energy resolution in addition to small size and high efficiency are crucial for medical physics applications such as positron emission topography (PET) and single photon emission computed topography (SPECT) and SiPM are now being considered for these applications [15,16].

The SensL[1] Series 1000 SiPM consists of an 848 pixel array of active area 0.43mm$^2$ arranged on a total device area of 1mm$^2$. Manufactured as an n on p device, the n implantation layer is very shallow, approximately 100nm, and the total depletion depth is of order 1μm. The device was supplied for evaluation without a protective can, the bond-wires secured in place with transparent resin. The main characteristics of the device are listed in table 1.

Table 1. Characteristics of SensL series 1000 SiPM device and SensL 2000 preamplifier.

| Pixel size | Number of cells | Geometric efficiency % | Maximum gain | Risetime (ns) | Decaytime (ns) | Spectral range (nm) | $V_{breakdown}$ at 25°C |
|---|---|---|---|---|---|---|---|
| 20μm | 848 | 43 | 8×10$^6$ | 10ns | 100ns | 400 - 700 | 28.2V |

Each pixel operates in limited Geiger mode in which a discharge is initiated by a hole or photoelectron in the high electric field region of the p-n junction. Each pixel is connected in parallel to a common load with the effect that the output is the sum of all pixel signals. The SiPM device is operated with a reverse bias voltage typically 10-15% above the breakdown voltage, the pixels electrically decoupled from one another by individual polysilicon resistors (R) of several hundred kOhms. This over-voltage is defined by equation 1. The individual pixel capacitance is approximately 50fF and the charge accumulated in each pixel given by equation 2 is therefore typically 150fC. This yields a signal of several millivolts per pixel on a 50Ω impedance, the summed output signal passing through an external preamplifier. The gain has a linear relationship with over-voltage given by equation 3 where e is the charge of the electron. Following a pixel discharge the avalanche current produces a voltage drop at the resistor leading to a decrease in the local electric field and the discharge subsequently quenches. After a dead time, determined by the pixel resistance and capacitance as shown in equation 4, the electric field is restored and the pixel becomes reactivated.

$$\Delta V_{over-voltage} = V_{bias} - V_{breakdown} \qquad (1)$$

---

[1] SensL (www.sensl.com) specialises in the development of low light sensing silicon photomultipliers in addition to providing a full range of signal preamplifiers and data acquisition electronics.

$$Q_{pixel} = C_{pixel}(\Delta V_{over-voltage}) \qquad (2)$$

$$Gain = \frac{C_{pixel} \cdot \Delta V_{over-voltage}}{e} \qquad (3)$$

$$\tau_{recovery} = R.C_{pixel} \qquad (4)$$

Due to the small width of the depletion region and the short Geiger discharge SiPMs typically have a single photoelectron timing resolution of several hundred picoseconds FWHM [5]. It is therefore the case for the majority of applications that the timing properties of the output signal are representative of the type and duration of the light source, and not of the device itself.

The quantum efficiency (QE) is defined as the average number of electron–hole pairs created by the interaction of one photon in the depleted layer of the SiPM device. Although dependent on the wavelength of the photon this value is typically between 50 and 80%. The breakdown probability represents the likelihood that a single photoelectron will trigger a Geiger discharge within the pixel, and is strongly dependent on the electric field strength in the junction, and therefore on the bias voltage [17]. The geometrical efficiency represents the ratio of the total active pixel area to the entire device area. Dependent on the design of the inter pixel boundaries which inhibit current flow between pixels, this is typically between 30% and 70%. The photon detection efficiency (PDE) is defined in equation 5 and 6 where $\varepsilon_{Geiger}$ is the Geiger discharge probability and F is the geometrical efficiency ratio. Typical values of between 20 and 30% for the PDE are very similar to the QE of a standard quartz PMT.

$$PDE(\lambda)\% = \frac{generated\ photoelectrons}{incident\ photons\ (\lambda)} \qquad (5)$$

$$PDE(\lambda)\% = \varepsilon_{Geiger} \times QE(\lambda)\% \times F \qquad (6)$$

SiPM devices have been widely evaluated for room temperature application. However their performance at lower temperatures has not been fully characterized. To assess the suitability of the SiPM device for light readout applications within liquid noble gas targets, its performance must first be studied in isolation. The key aim of this work was to confirm cryogenic operation at liquid argon temperatures. Gain in LAAPDs is known to vary with temperature [10] by as much as 3%/°C and temperature dependence of gain has also been observed during evaluation of SiPM devices for room temperature operation within the ND280 of the T2K detector [18].

128nm VUV scintillation light produced within argon targets is typically shifted via chemical waveshifters into the sensitive spectral region of the readout device [19]. Traditionally the QE of PMTs reaches a maximum at approximately 430nm. Since the exact emission maxima can be adjusted via multi-waveshifter blending, it is appropriate that the QE of the SiPM device also be investigated over a number of photon wavelengths at low temperature. Any deterioration in performance at low temperature and at high photon fluxes caused by saturation effects must also be quantified. This work details these measurements.

## 2. Experimental Details

The apparatus, shown in figure 1, used in all measurements consisted of an inner steel test chamber of height 550mm and diameter 98mm held within a concentric outer chamber of height 750mm and diameter 250mm. Both chambers were contained within a copper vacuum vessel and access to each chamber was provided through tubes in an exterior top flange of diameter 420mm. The 1mm$^2$ SiPM device was positioned near the centre of the inner chamber hanging from an optical fibre cable as shown in figure 2. A pulsing circuit based on that of Kapustinsky [20] was connected to an externally mounted light emitting diode (LED) producing 5ns rise-time 15ns decay-time light pulses of variable intensity, which were passed through the fibre optic cable to the SiPM device. The output of the optical fibre was positioned less than 1mm from the SiPM device so that the total area was illuminated. The fibre was then fixed to the support structure to ensure that its position relative to the device remained unchanged despite thermal contractions. The preamplifier was powered using a Digimess DC power supply HV3003-2 and the bias voltage supplied by a Thurlby Thandor PL320 32V, 2A DC supply with sense active. The output signal, viewed on a Lecroy 9350A scope prior to each measurement, was passed to an Acqiris CC108 PCI acquisition system typically triggered directly from the LED generator gated logic signal ensuring a trigger on every light pulse. PT1000 platinum resistance thermocouples attached to the SiPM provided accurate temperature information, and careful addition of liquid nitrogen to the cryogenic jacket enabled the temperature of the SiPM to be stabilized for any time to an accuracy of ±1°C. A parallel plate capacitor was placed above the device to act as a level sensor, indicated by the change in the dielectric medium. The only thermal inputs to the SiPM device were radiation from the walls and conduction along the support structure and power lines. Prior to all tests the inner chamber was evacuated to $1 \times 10^{-7}$ mbar and backfilled with gaseous nitrogen to 1bar to avoid condensation of water or other gases in the optical pathway. For all measurements as the temperature decreased, the pressure of nitrogen gas within the target was maintained at 1 bar until liquefaction eventually occurred.

Figure 1. Experimental apparatus, ancillary equipment, and data acquisition system.

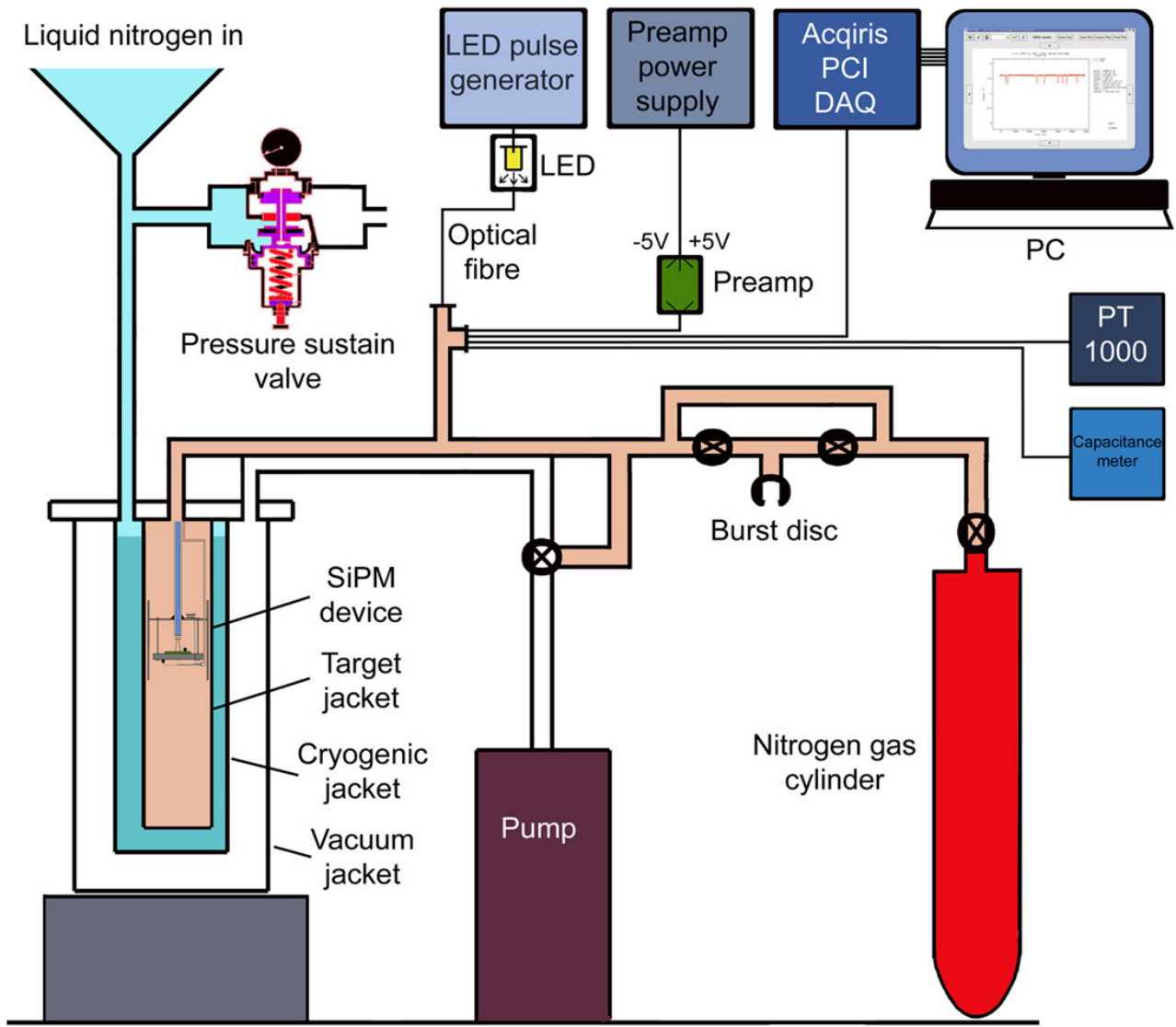

Figure 2. Internal structure used to evaluate cryogenic performance of SiPM device.

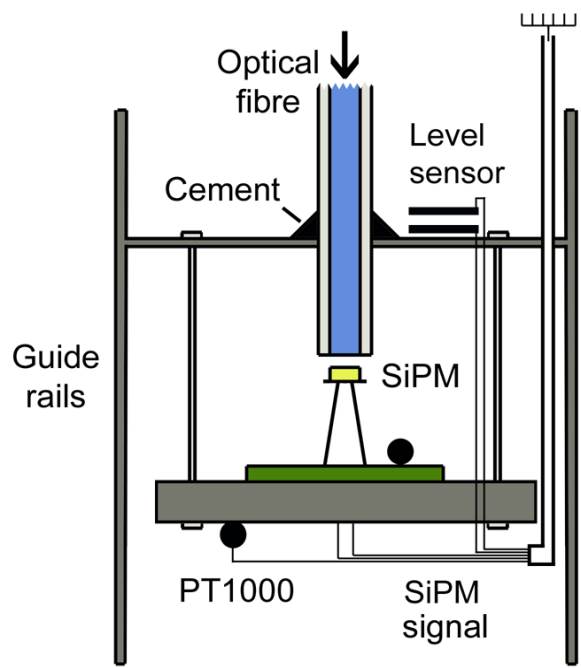

## 3. Results

To assess the performance of the series 1000 SiPM device for liquid argon applications, a number of key performance measurements were taken.

### 3.1 Low temperature limitations of the preamplifier

The SiPM device was attached directly to a high gain pulse preamplifier printed circuit board comprised of a large number of resistors and capacitors. Since this was initially mounted within the cryostat, preliminary measurements were undertaken to ensure that the performance of the preamplifier was unaffected by cryogenic temperatures. The SiPM device was removed and replaced by an RC circuit (R=50Ω, C=12.4pF at 25°C and C=12.5pF at -196°C) connected in parallel to a signal generator supplying a 200Hz square wave pulse of variable amplitude. Amplifier clipping, output signal distortion, and gain were then measured as the preamplifier was slowly cooled.

At room temperature preamplifier clipping was observed at 1.5V output signal height for a supply voltage of ± 5V. As the temperature was reduced for a constant signal generator input voltage, the gain was also noted to decrease, as shown in figure 3. Distortion of the output signal was also observed for output pulse heights above 0.7V at -150°C reducing to pulses above 0.1V at -196°C, although for both room temperature and liquid nitrogen operation there remained a proportional albeit reduced linear relationship between the input and output voltages, so long as the distortion threshold was not exceeded. The preamplifier was therefore considered inappropriate for use in liquid argon environments and removed from the cryostat for all tests. Careful screening of the supply cables between device and preamplifier was found to be crucial to reduce noise. This was achieved using an ultrahigh vacuum Kapton insulated coaxial cable, surrounded by an additional grounded steel braid, the connections passing out of the target chamber through two BNC welded fittings.

Finally by comparing the area of the pulses measured by the data acquisition system with the charge deposited at the input of the pre-amplifier (knowing $V_{square}$, C and R), a calibration of the device gain was made, allowing the absolute gain to be determined from the separation of the photoelectron peak maxima.

Figure 3. Preamplifier performance limitations imposed by low temperature operation.

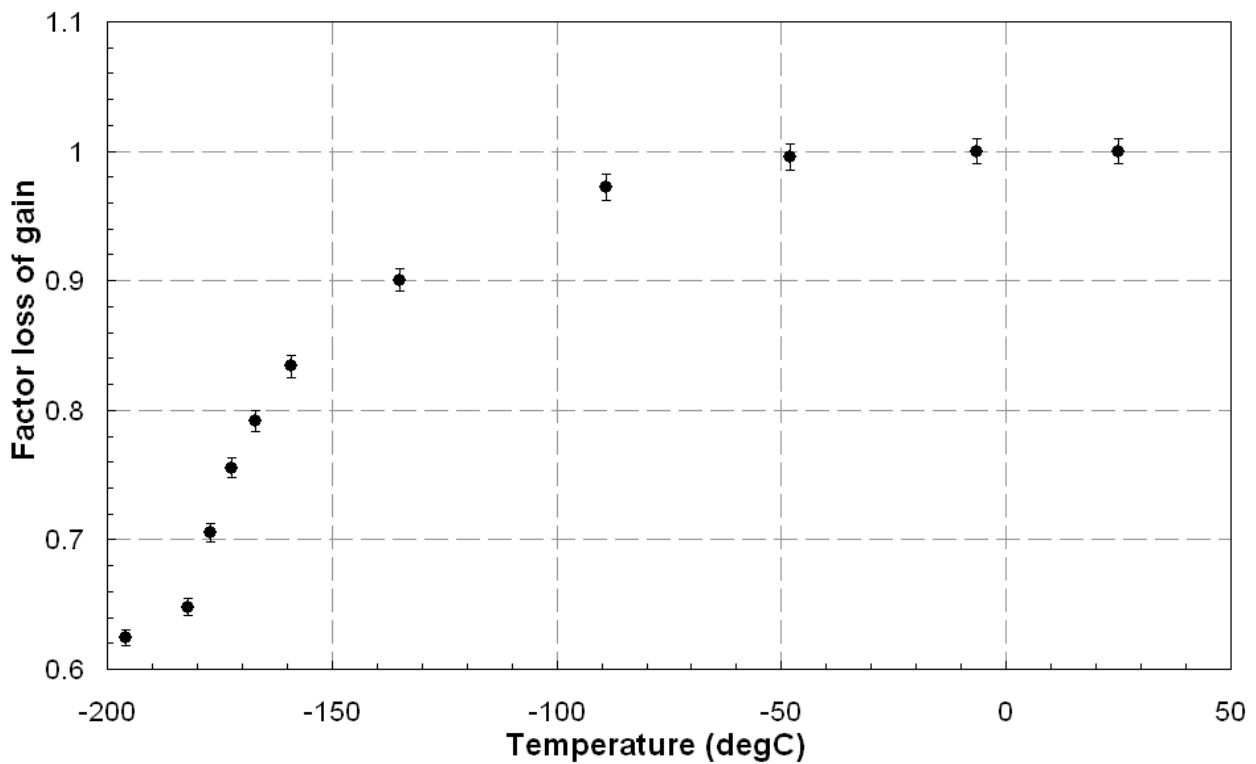

## 3.2 Variation of gain and breakdown voltage with temperature of SiPM device

The effect of cooling the SPM1000 SensL device slowly to -196°C on the breakdown voltage and the gain was assessed. Very low intensities of 460nm blue LED pulses were selected so that the individual photoelectron spectrum could be more easily observed, the separation of the photoelectron peaks shown in figure 4 being the standard method by which the gain is measured as detailed above.

Figure 4. Single photoelectron peak distributions at 25°C (unshaded) and -196°C (shaded) at bias voltages of 31.5V and 27.5V respectively.

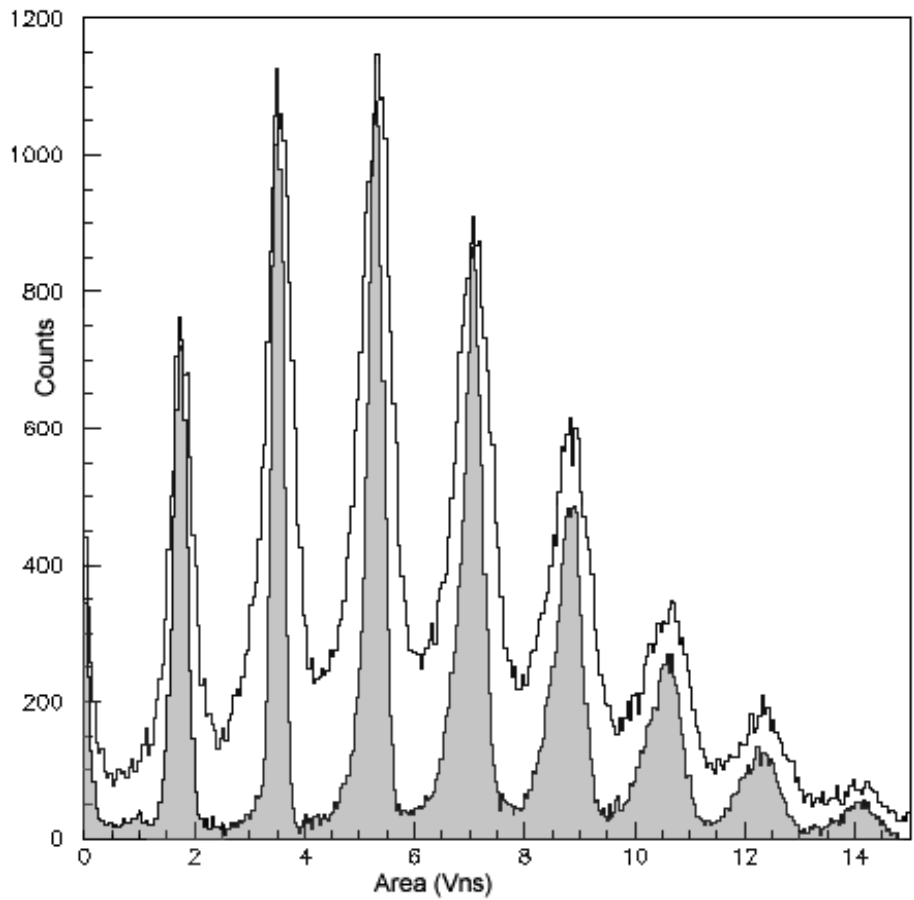

Measurements taken within the apparatus at room temperature confirmed the device was operating correctly, SensL having previously measured a nominal $V_{breakdown}$ at 25°C of 28.2V. Liquid nitrogen was slowly added to the cryogenic jacket to produce a 1°C per minute cooling rate until the temperature of -196°C was reached. At this point nitrogen gas within the target vessel began to condense and this continued until the level sensor indicated the SiPM device was fully immersed. During cooling, the photoelectron spectra were recorded for a range of $V_{bias}$ values and temperatures. On completion the liquid nitrogen in the cryogenic jacket was completely removed by pressurisation, and the liquid nitrogen in the target allowed to slowly vent. During this warming, all measurements were repeated to test for hysteresis, none being observed.

Excellent single photoelectron resolution for both single photon counting measurements and moderate light intensity LED pulses indicate a high level of performance uniformity amongst the pixels allowing each photoelectron peak to be clearly distinguished and fitted using a Gaussian function. The device gain measured as a function of bias voltage for a range of temperatures is shown in figure 5.

Figure 5. Dependency of the gain on the bias voltage for a range of device temperatures.

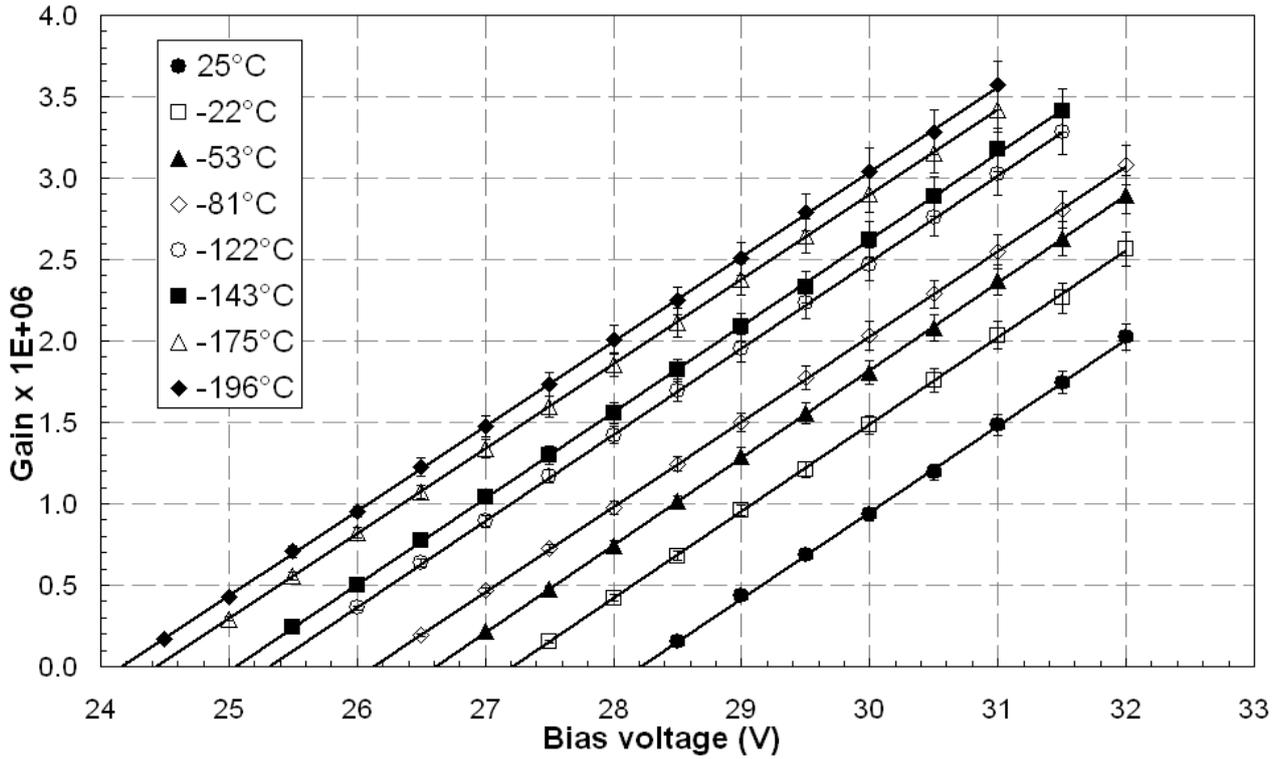

The breakdown voltage has long been known to be heavily dependent on temperature. A linear fit was made to the data points in figure 5 allowing $V_{breakdown}$ (the bias voltage at zero gain) to be measured. This is displayed in figure 6 plotted against temperature, the drop in the $V_{breakdown}$ coefficient measured as 18.3±2.5mV/°C.

Figure 6. Dependency of the breakdown voltage on device temperature.

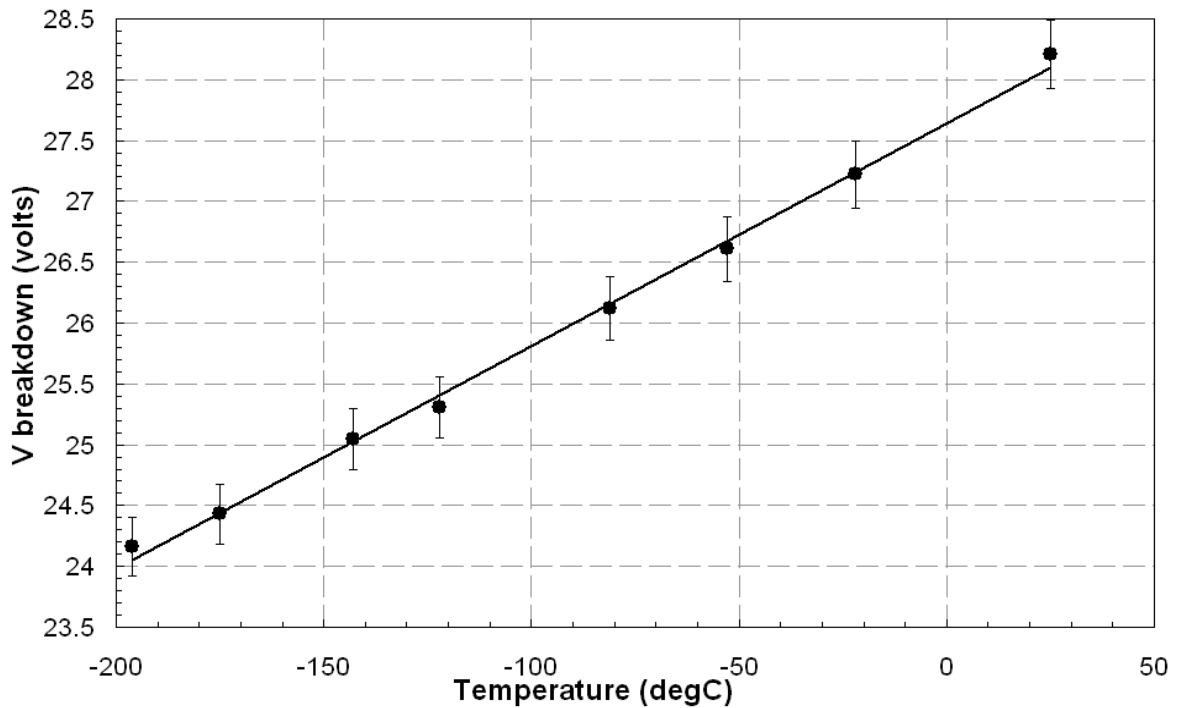

The over-voltage $\Delta V$, defined in equation 1 can then be calculated for any temperature, and the gain measured for specific values of $\Delta V$ as shown in figure 7.

Figure 7. Gain as a function of temperature for constant over-voltages, ΔV.

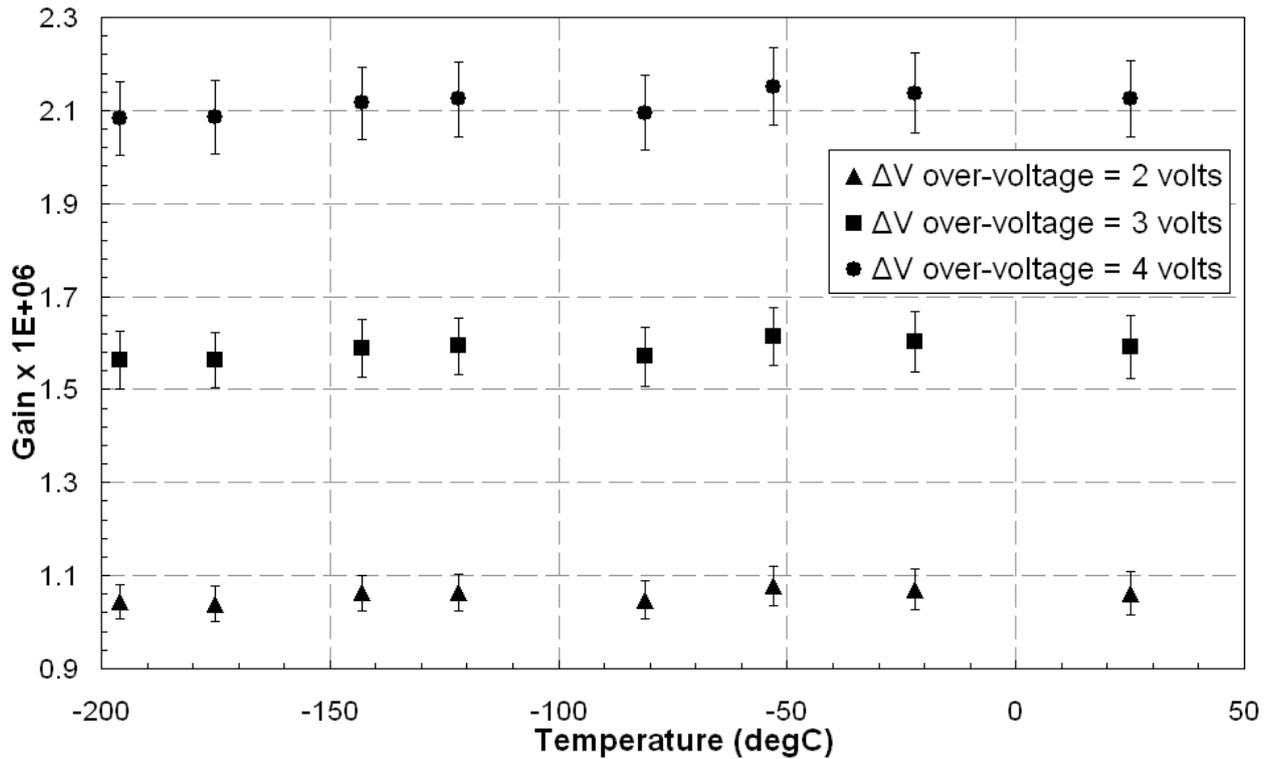

All data indicates stable linear device gain for a constant over-voltage from room temperature to typical liquid argon temperatures of -190°C.

3.3 Spectral response and PDE of the SiPM device as a function of temperature

The spectral response and PDE of the device were evaluated using small pulses from 460nm blue and 680nm red LEDs. Typically the LEDs were pulsed at a frequency of 100Hz, each pulse lasting 20ns and typically containing between 10 and 1000 photons. The response of the SiPM to the LED pulse was defined by the number of photoelectrons produced in the device due to the incidence of one light pulse on the pixel array, the number of pixels fired referred to as the number of photoelectrons within the output signal. The absolute number of photons within each incident pulse was measured using a 2 inch Electron Tubes D749QKFLA low temperature PMT of known QE and gain dependency with temperature. In order to eliminate solid angle effects the quartz PMT window was covered in highly reflective 3M foil apart from a small central $1mm^2$ hole, and the SiPM and fibre optic cable were also wrapped in 3M foil. The use of an optical fibre deteriorates the timing resolution of the LED pulses, which should be certainly no more than the approximate pixel recovery time. The transmission properties of the optical fibre were therefore measured as a function of decreasing temperature using the calibrated low temperature PMT. No variation in the measured intensity of a pulsed LED signal was observed for the short distances used.

The LED pulse frequency was held constant at 100Hz and the intensity adjusted to be below the saturation threshold for either the preamplifier or the SiPM device. For each LED, the number of photoelectrons created was then compared with the initial number of photons incident on the SiPM device, over a range of temperatures and over-voltages. The mean number of photoelectrons observed was determined by fitting to the data to obtain a mean and standard deviation. An example plot is shown in figure 8. For high light intensities the observed peak was clearly defined by multiple photoelectrons as shown in figure 15(b). At low intensities, such as that shown in figure 4, the peak was less well defined.

Figure 8. Number of photoelectrons generated due to 460nm incident photons for four over-voltages at -196°C.

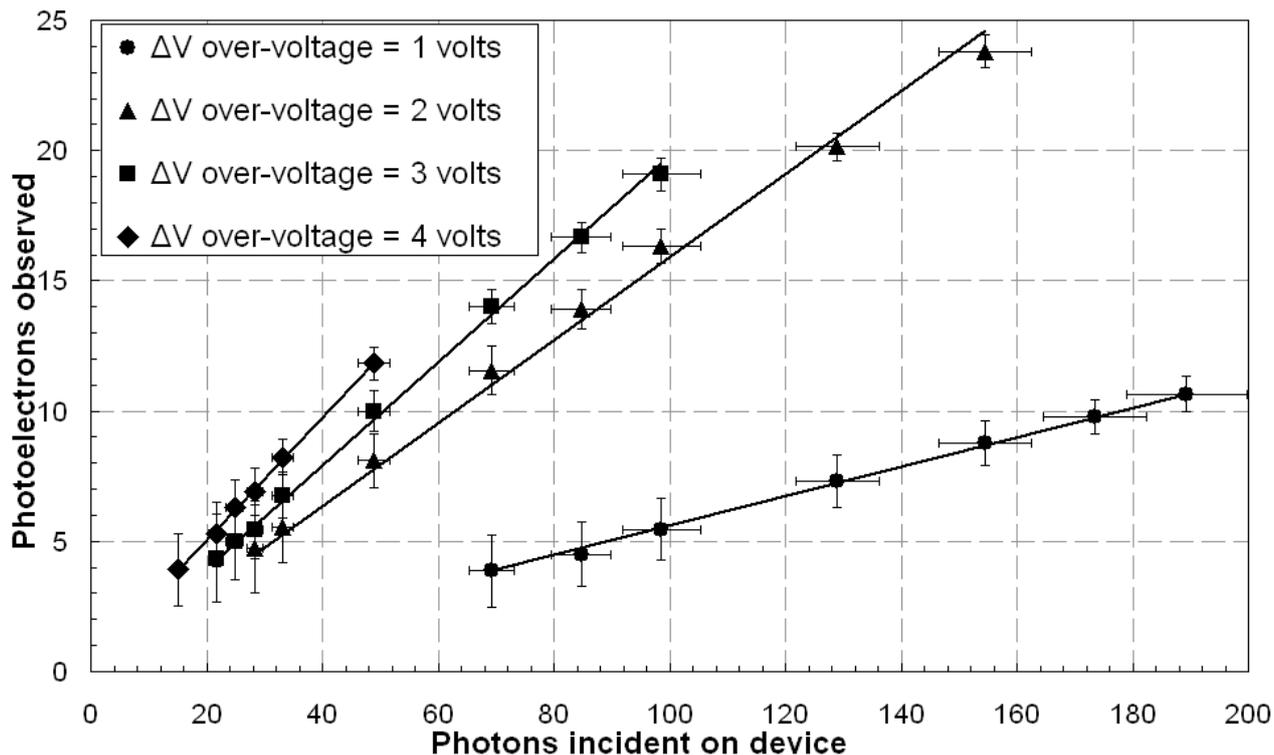

Extensive measurements revealed that although the device PDE was sensitive to the over-voltage, it was found to be insensitive to temperature within errors so long as the bias voltage was reduced to maintain a constant over-voltage with decreasing temperature. The PDE, calculated using equation 5 for a range of over-voltages and temperatures, is shown in figures 9 and 10 for both 460nm and 680nm photon LED 100Hz pulses. It should be noted that a small proportion of the observed signal in any measurement is due to crosstalk and afterpulsing.

Figure 9. PDE as a function of over-voltage for 460nm and 680nm pulsed light at both 25°C and -196°C.

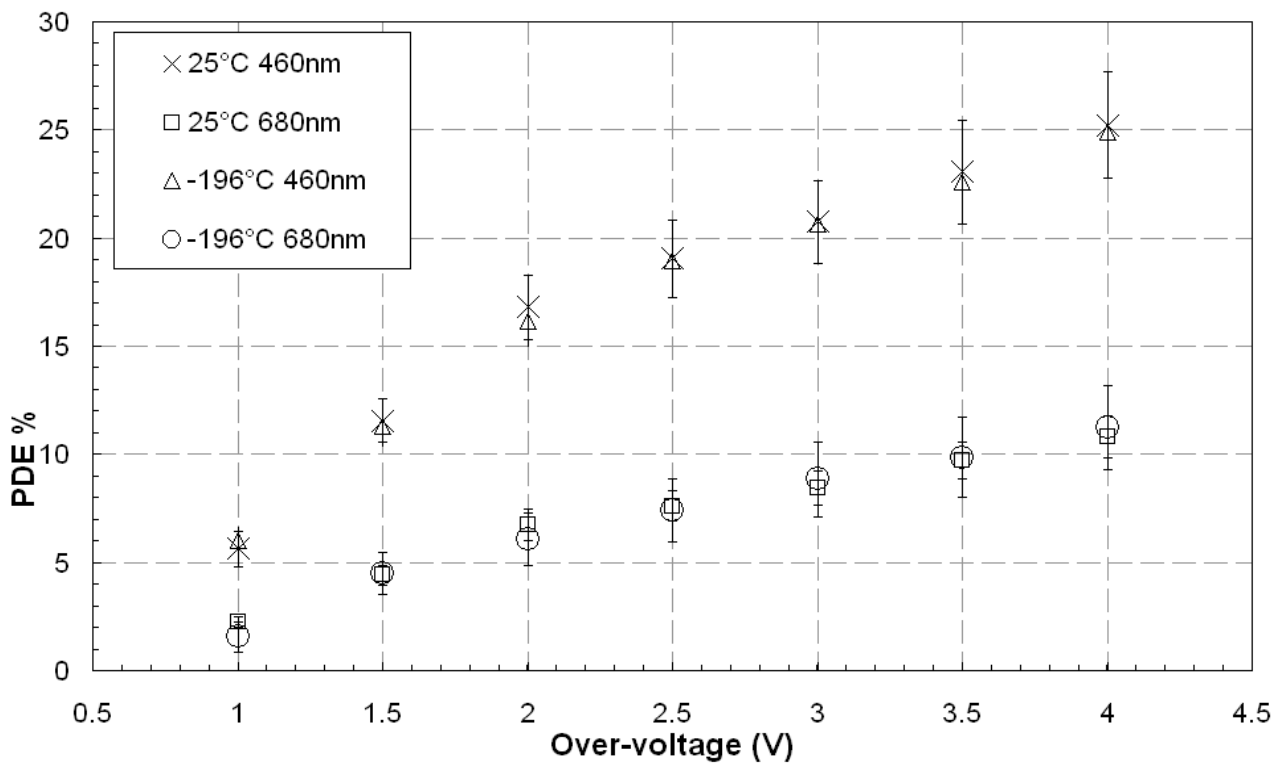

Figure 10. PDE as a function of temperature for 460nm and 680nm pulsed light at three over-voltages.

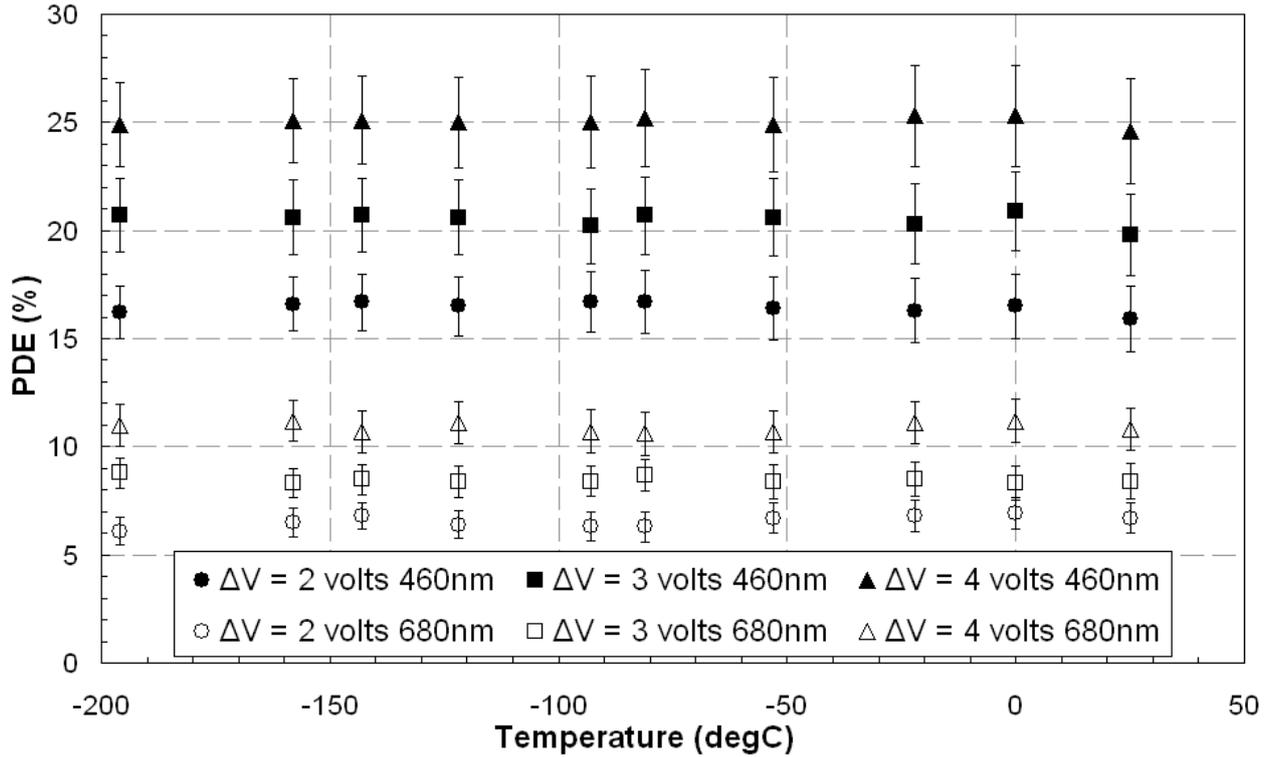

In order to both identify any hysteresis and recheck data, a select number of measurements were repeated as the device was gradually warmed at the conclusion of each test. No variation in any measurement was observed outside errors.

3.4 Effect of temperature on the dark count rate

The dark count rate was measured as a function of both bias voltage and temperature. The LED pulser was switched off and the optical fibre thoroughly isolated from light. A signal generator producing a 10 kHz square wave gate was connected to the input trigger of the PCI acquisition system. The acquisition window was set at 150ns and data was recorded at fixed temperatures over a range of bias voltages. Since the PCI system was triggered by the signal generator and not the SiPM output signal or the LED pulser unit, a high proportion of events contained only noise, whilst a fraction contained a single photoelectron pulse, and a much smaller fraction contained multiple pulses. Events also featured truncated pulses at either the start or the end of the acquisition window. Although the noise pedestal, distributed approximately about zero area was the predominant signal, the data also identified maxima due to single photoelectrons and double photoelectrons. By fitting Gaussians to both the pedestal noise and photoelectron distributions, measuring the position of the peaks in terms of area, and then finding the midpoint between them, the total number of counts above this level could be determined. The dark count rate was then defined as:-

$$Dark\ count\ rate_{0.5pe}(Hz) = \frac{total\ number\ of\ counts\ above\ 0.5\ photoelectron\ area}{total\ number\ of\ counts\ taken \times 150ns\ window} \quad (7)$$

$$Dark\ count\ rate_{1.5pe}(Hz) = \frac{total\ number\ of\ counts\ above\ 1.5\ photoelectron\ area}{total\ number\ of\ counts\ taken \times 150ns\ window} \quad (8)$$

The dark count rate above 0.5 photoelectrons is shown in figure 11. Although $5 \times 10^6$ events were not absolutely necessary for room temperature data, the dark count rate was found to drop exponentially with decreasing temperature, and even 5,000,000 events yielded poor albeit acceptable statistics for the first photoelectron distribution at -196°C.

Figure 11. Dark count rate variation with bias voltage and temperature of the SiPM device.

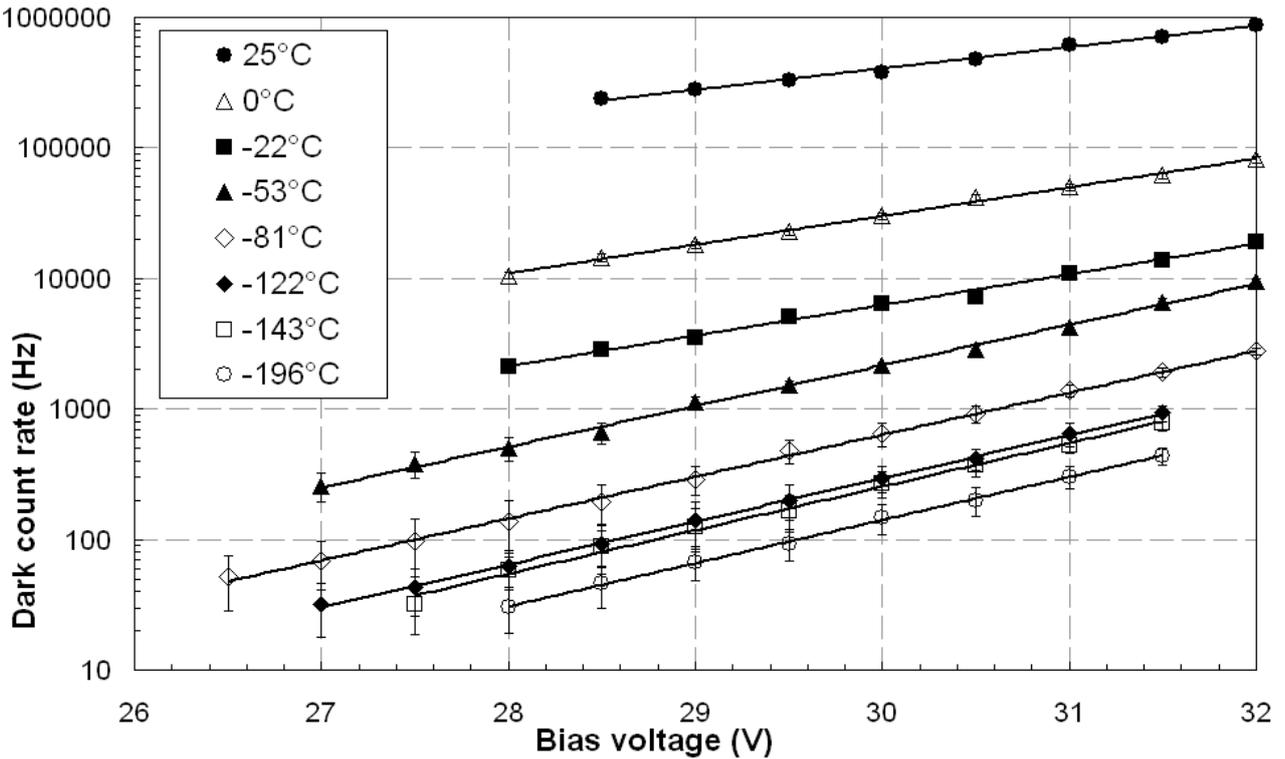

Figure 11 illustrates the exponential increase in the dark count rate with bias voltage for a constant temperature, suggested as due to electric field assisted tunnelling.

Since the breakdown voltage drops with decreasing temperature, and since the temperatures for which data were taken coincided with previous measurements, the exponential fits to the curves in figure 11 were used to provide a measure of the variation in the dark count rate as a function of temperature at constant over-voltages. This data is shown in figure 12.

Figure 12. Dark count rate as a function of temperature for three over-voltages.

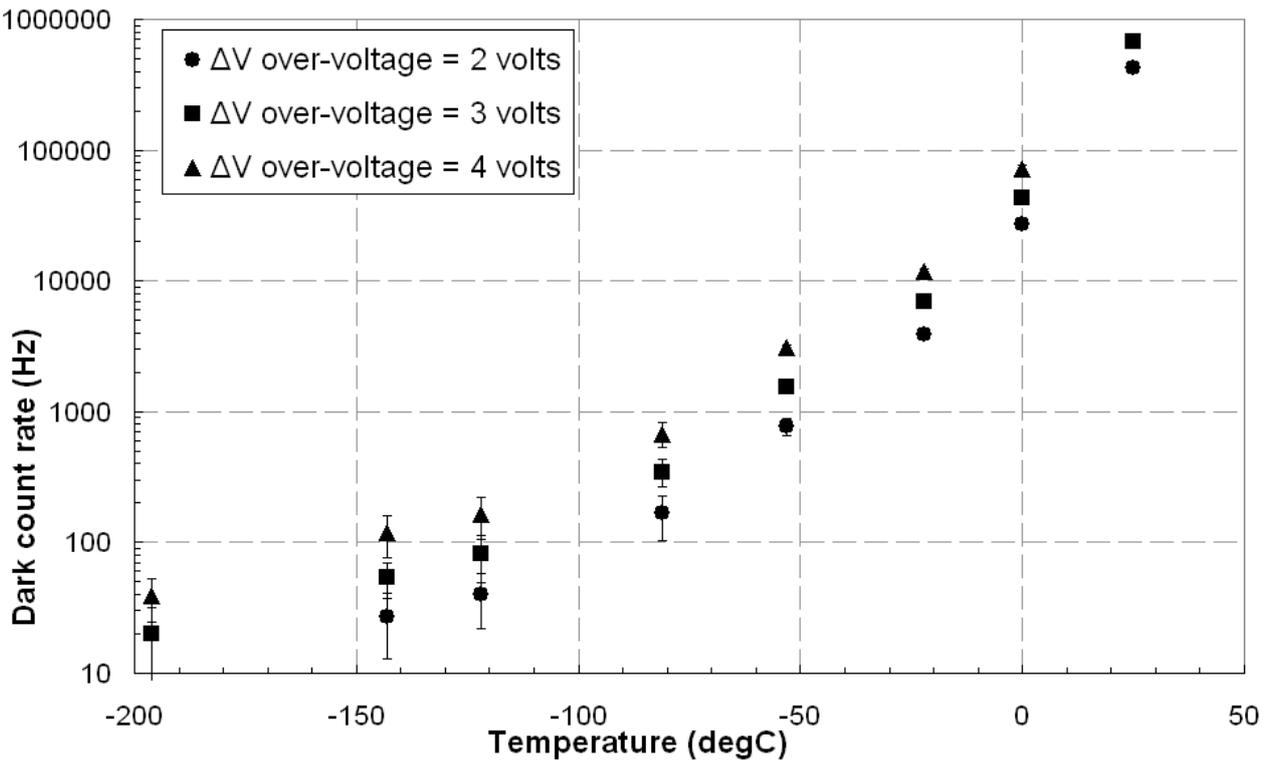

Finally an attempt was made to identify the contribution to the total dark count rate due to optical crosstalk. Crosstalk and afterpulsing can lead to an overestimation of the PDE as the Geiger discharge occurs in additional pixels. If it is assumed, for dark count rates significantly smaller than the DAQ integrating time window, that dark noise in which more than one pixel is fired comes predominantly from a combination of dark noise and crosstalk with neighbouring pixels, the probability for crosstalk can be given by equation 9 [21,22].

$$\text{Probability of crosstalk} = \frac{\text{rate of} \geq 2 \text{ pixel fired noise}}{\text{rate of} \geq 1 \text{ pixel fired noise}} \qquad (9)$$

The device was allowed to warm to room temperature, slowly cooled and the crosstalk probability calculated for a range of over-voltages for a number of specific temperatures as shown in figure 13. Unfortunately despite each data set containing $5 \times 10^6$ events, as the temperature reduced, the total dark count rate quickly dropped and the crosstalk probability ratio became limited by low statistics for all temperatures below -53°C. However up to this point the crosstalk probability remained independent of temperature as shown in figure 13.

Figure 13. Crosstalk probability as a function of over-voltage over a range of temperatures.

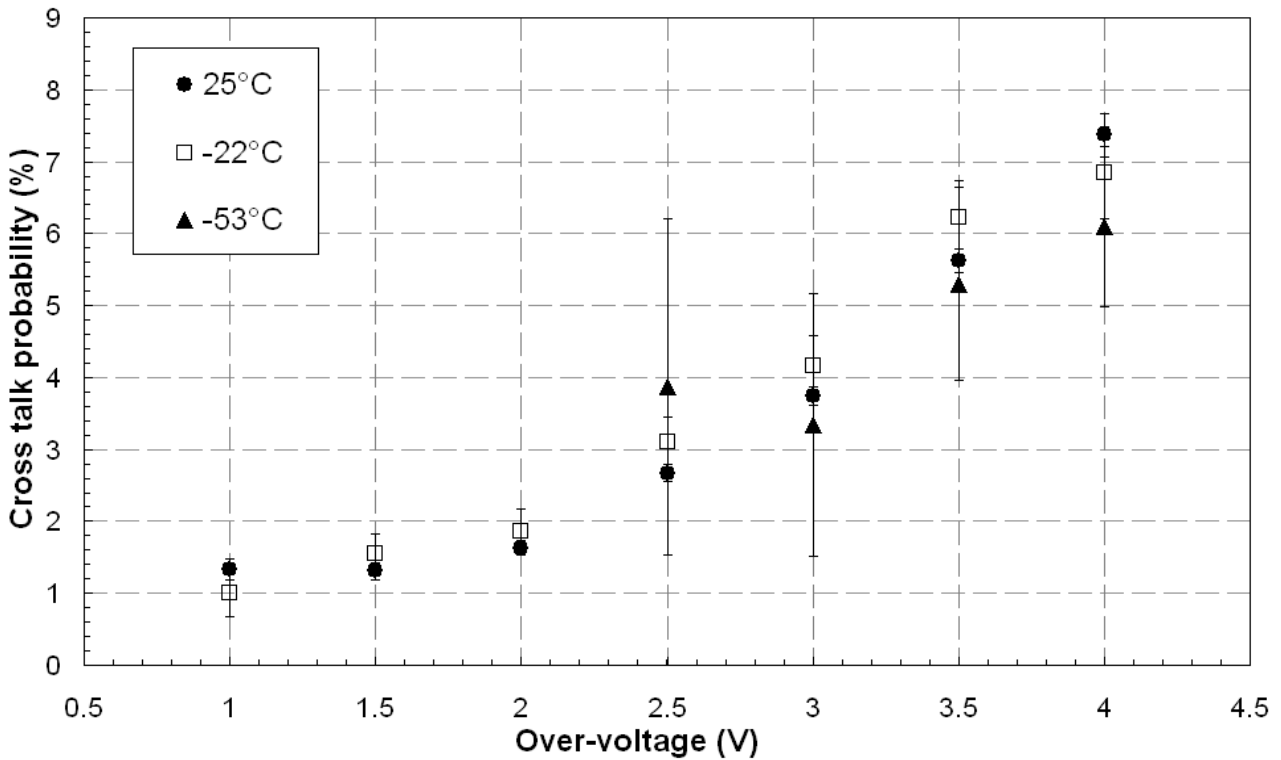

### 3.5 Effect of temperature on SiPM device saturation due to high LED photon flux

An evaluation of the linearity of the output signal from the SiPM with increasing light intensity as a function of temperature was made. Saturation of the SiPM device was considered to be a function of the bias voltage, the light intensity, the pulse frequency, and temperature. In all tests the pulse frequency was maintained at 100Hz. For a fixed temperature and over-voltage, the intensity of the 460nm LED pulse was gradually increased from 150 to 1600 photons, the LED flux calibrated using the ETL D749QKFLA PMT as described earlier, until either the saturation point was reached or the output signal became clipped by the preamplifier. The bias voltage was then increased and the procedure repeated. This procedure was then repeated for lower temperatures.

Earlier measurements, illustrated in figure 8, have shown linearity of the output signal is maintained at low incident photon fluxes irrespective of the over-voltage or temperature. Unfortunately due to a combination of the single photoelectron height, the 1.5V preamplifier clipping level, and the increased PDE for 460nm light at high over-voltages, no deviation from proportionality could be observed in any measurement at any temperature for any over-voltage above 1V, the preamp saturating at low photon fluxes. Data taken at room temperature and -196°C and shown in figure 14 indicates deviation from linearity to be both independent of temperature, and to occur for 1V over-voltages for incident fluxes containing in excess of 300 photons. On cessation of the photon flux at the conclusion of each measurement the device was noted to return to its typical dark count background rate within less than one second.

Figure 14. Effect of temperature on SiPM device saturation due to high LED photon flux.

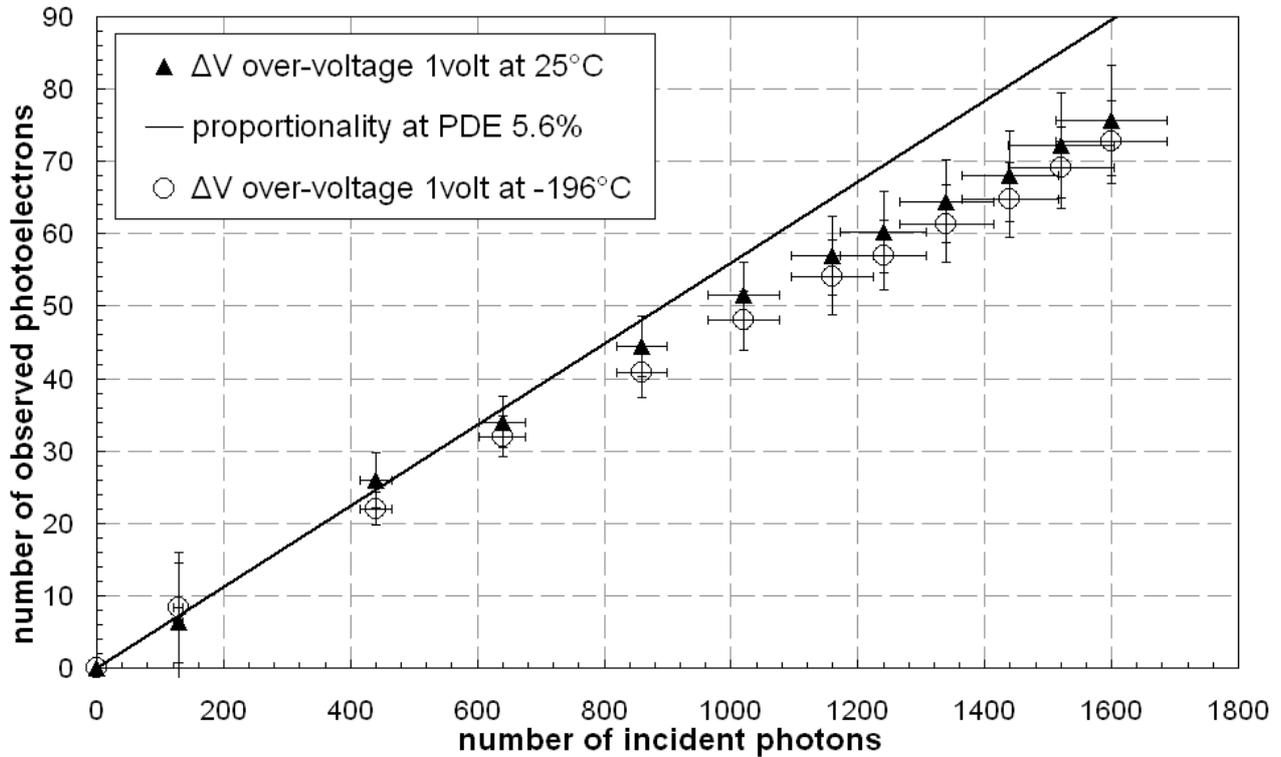

3.6 Effect of cryogenic operation on the single photoelectron pulse shape

During a Geiger discharge within a single pixel, lateral diffusion of free carriers assisted by avalanche multiplication [23] spreads the development of the discharge across the pixel surface. The rate of discharge development may therefore be a function of temperature. The rise-time, decay-time, and standard deviation of both single photoelectron and LED pulses were measured as a function of temperature, the over-voltage held constant throughout at 1V.

Mean single photoelectron rise-time and decay-times of 9±2ns and 105±12ns respectively were recorded at room temperature. The mean pulse area and corresponding standard deviation were also measured to be 0.51±0.15Vns as shown in figure 15(a). As the temperature of the device was lowered below -150°C the shape of the single photoelectron pulse was noted to change. Although the total pulse area and the rise-time remained unaffected, the decay-time increased to 130±12ns. The mean single photoelectron pulse area at -196°C was 0.49±0.07Vns, the bias voltage having been reduced to 25.2V to maintain a constant over-voltage of 1V.

Measurements were then taken both at room temperature and -196°C using a 460nm LED producing 100Hz pulses containing a varied number of photons in order to characterize any performance limitations at high light fluxes. So as to assess any additional impact on the rise-time, decay-time and standard deviation of the LED pulse due to saturation effects for photon fluxes in excess of the limit of proportionality identified in section 3.5, data was taken for incident fluxes up to 2000 photons. No effects were noted as the photon flux was gradually increased from 300 to 2000 photons per pulse. No changes inconsistent with single photoelectron behaviour were observed either for low photon fluxes or for high fluxes above the saturation level, the mean pulse area and corresponding standard deviation of a pulse containing 1440 photons measured as 33.1±0.8Vns at 25°C and 32.9±0.48Vns at -196°C and shown in figure 15(b).

Figure 15. Comparison of (a) single photoelectron resolution and (b) LED pulse resolution at 1V over-voltage: (unshaded) 25°C; (shaded) -196°C.

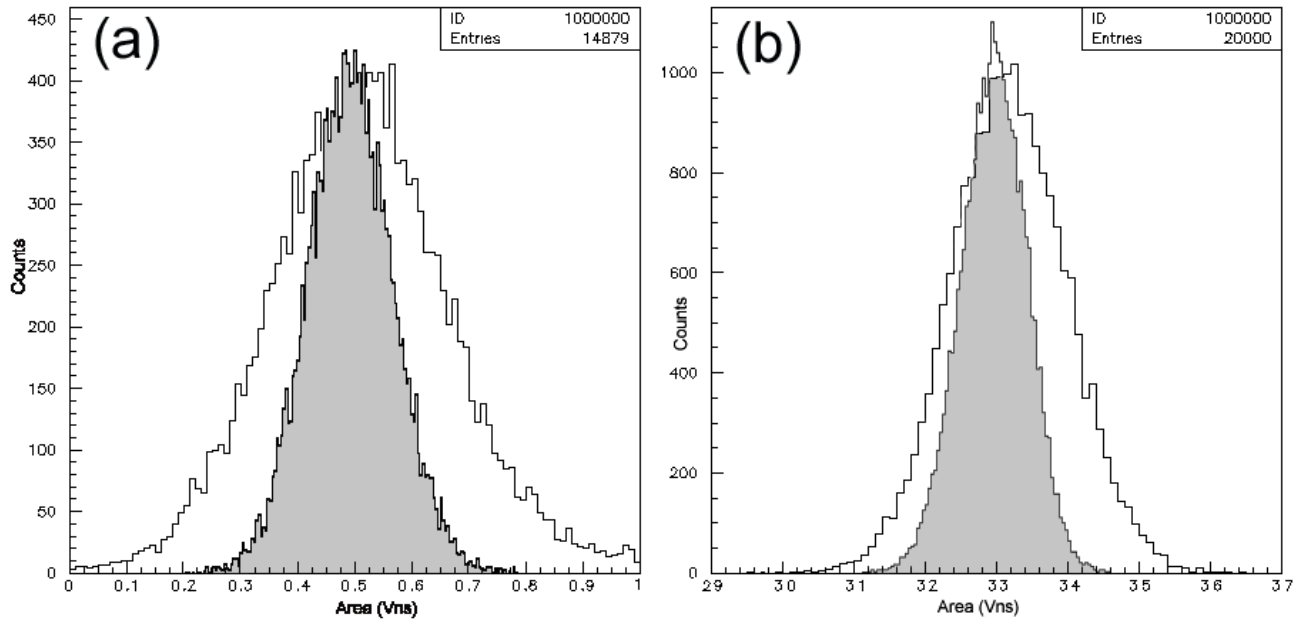

3.7 Resilience of the device to repetitive thermal shock

In the course of this investigation the SiPM device was thermally cycled many times and exposed to both sustained high bias voltages and light intensities. In order to assess any lowering of performance due to ageing or physical damage, the device assembly including the optical fibre was removed from the test chamber and installed in a small metal box, a hole drilled through the bottom and top walls. The box was then immersed directly in liquid nitrogen, and once cryogenic temperature had been achieved, removed and the liquid nitrogen drained off. This process was repeated 30 times over a 6 hour period, the device allowed to warm to room temperature prior to being re-immersed. Measurements of the gain and dark count rate were taken at three over-voltages at every fifth immersion, shown in figure 16. The resulting data was within errors equal to the initial measurements of gain taken in section 3.2 indicating no physical damage had been done to the device.

Figure 16. Device gain measured at -196°C for three over-voltages as a function of the number of thermal cycles carried out on the SiPM device.

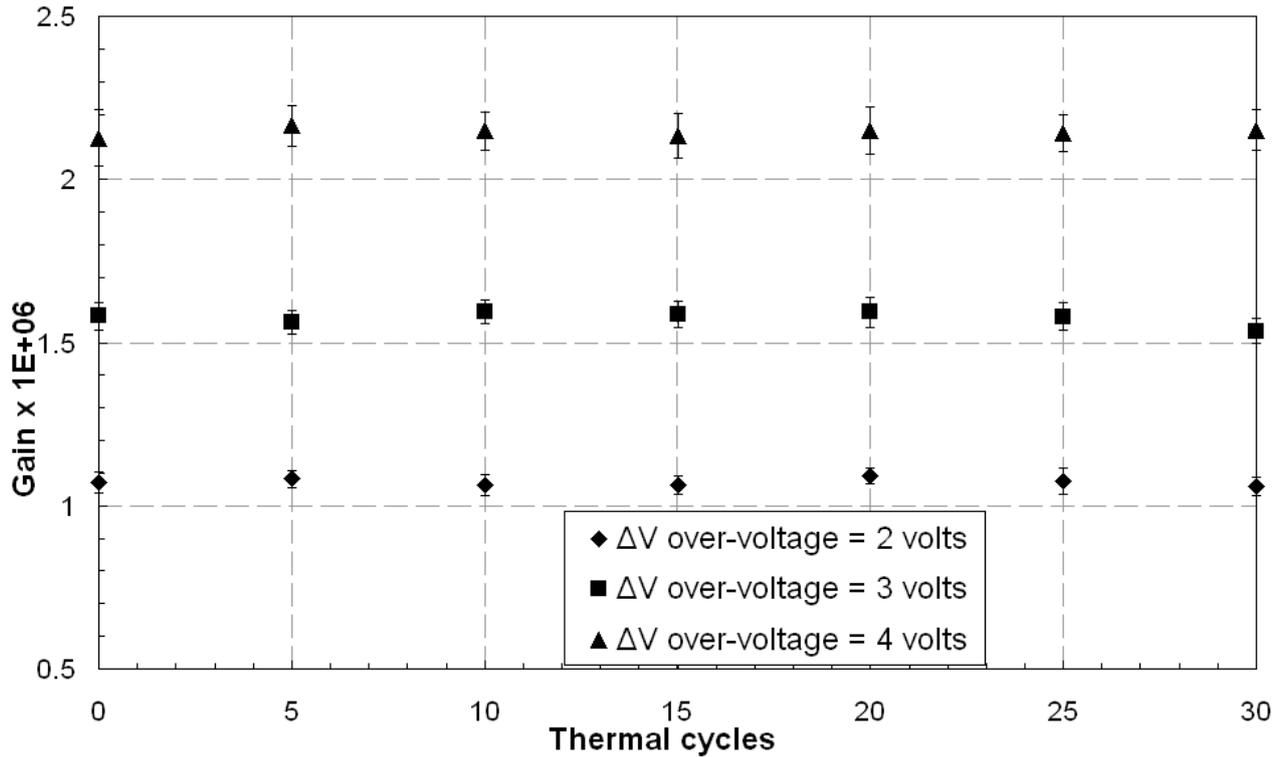

## 4. Discussion

Results obtained indicate that the low temperature gain of the SensL SiPM device is well suited for light detection in cryogenic environments, its photon detection efficiency at 460nm is well matched to liquid argon targets featuring the waveshifter tetraphenyl butadiene (TPB), and in general is a good match for medium and high energy physics applications such as neutrino physics and dark matter detection. PDE, dark count rate, and single photoelectron timing are comparable with traditional quartz PMTs, and the dynamic range has been shown to be linear for typical photon fluxes per solid angle produced in liquid argon TPCs although this must be confirmed using simulations which should include purity, drift distance, etc.

PiN photodiodes, Geiger mode avalanche photodiodes, and silicon photomultipliers have already replaced traditional quartz PMTs in a number of applications. LAAPDs have been extensively used to measure scintillation photons within for example the Compact Muon Solenoid (CMS) [24] and Positron Emission Topography (PET) scanners [25]. The T2K (Tokai to Kamioka) long baseline neutrino oscillation experiment [26], due to take first data in 2009 has selected SiPM devices as the baseline photosensor for the ND280 near detector ECAL, and their use is also being considered in the calorimeter of the International Linear Collider.

Unlike quartz PMTs, SiPM device performance can be heavily optimised for a specific purpose during both the manufacturing stage and in normal operation. As can be seen, a multi-dependency exists between the over-voltage, the PDE, the dark count rate, and the gain, although pixel size and dynamic range, device layer structure and pixel recovery time also play a role. Inevitably a compromise must be reached between all variables for each specific application.

The optimum over-voltage for peak signal to noise ratio is generally a compromise between the PDE and the dark count rate, both of which increase with voltage. This is particularly relevant for applications, such as liquid argon TPCs, requiring a long integration time. Operation at liquid argon temperatures reduces the dark count rate by $1\times10^{-5}$ of its equivalent room temperature value, with the effect that the bias voltage can be increased, leading to greater device gain and PDE due to an enhanced Geiger discharge probability $\varepsilon_{Geiger}$ and ionisation coefficient. The PDE and gain at -196°C for 460nm photons were found to increase to 25% and $2.1\times10^{6}$ respectively for an over-voltage of 4V whilst the corresponding dark count rate above a discrimination threshold of 0.5 photoelectrons was only 40Hz. There is also scope to raise the over-voltage to a maximum of about 6V so further increasing both the gain and PDE.

The PDE of a SiPM device is mainly determined by the photon absorption length in silicon. Unlike long wavelength photons (>600nm) which penetrate deeply into the device and are mostly absorbed in the non depleted bulk, short wavelength photons (<400nm) are typically absorbed within 100nm of the silicon surface. It is therefore more likely that generation of an electron – hole pair occurs in the highly doped top implantation layer where short recombination times result in a loss of charge transfer. Manufacture of SiPM devices for which the PDE is maximum for blue and violet wavelengths therefore necessitates the use of extremely thin $p^{++}$ or $n^{++}$ top layers, in contrast to red and infra-red sensitive devices which require thicker depletion layers. Electrons are more likely to initiate a breakdown in silicon than holes [27], and therefore generation of free carriers in the p region has a higher trigger probability. Devices optimised for blue light generally consist of p-silicon on an n-substrate. SiPM devices have been manufactured at the Joint Institute for Nuclear Research (JINR) on a n-type substrate with a QE spectral response peaked at 460nm and 10,000 pixels/mm$^2$ at a gain of $3\times10^4$ [28]. Future devices may feature enhanced PDE by maximizing the quantum efficiency and transmittance of charge, whilst reducing the likelihood of recombination. This may be achieved by both optimizing the thickness of the silicon layer and reducing its impurity concentration. Reflections at both the protective outer polymeric coating and the $SiO_2$ surface can be reduced with the use of pure materials and the use of specialist non reflective coatings.

The dark count rate has been observed to decrease exponentially with temperature at a constant over-voltage. This implies that the dominant component is due to single pixel discharge initiated by thermally generated carriers, although field assisted generation of free carriers, which is temperature independent and can only be reduced by decreasing the bias voltage, also has an effect. Due to the high gain, the SiPM electronics noise is very small. The dark count rate has recently been shown to be almost proportional to the active area suggesting that the impact on the count rate of large area SiPM operation at cryogenic temperatures would be minimal [29].

Development of SiPM technology will inevitably be determined by market forces, with the effect that devices may not necessarily be fully optimised for specialist academic applications such as operation in liquid argon. Medical physics applications such as PET offer a far larger market to SiPM industry, in which large area SiPM devices, operating within a magnetic field, would be coupled to LSO or LYSO crystals emitting a high number of photons peaked in the blue region of the spectrum. Due to the high flux from 511keV X-rays, the dark count rate typically taken above a threshold of no less than 5 photoelectrons, would not be an issue, although timing resolution would be fundamental. These characteristics are an excellent fit to particle physics applications such as neutrino or dark matter detection.

The SensL series 1000 SiPM device has been demonstrated to have high gain with low electronic noise, excellent photon counting capabilities, predictable temperature and voltage stability, at low cost and in a compact high purity design with position sensitivity of 1mm$^2$. SiPMs are rugged, show no aging, and can tolerate accidental illumination. Although SiPM devices have not yet achieved the performance parameters of traditional quartz PMTs especially in terms of active area and high rate capability, development of SiPMs is ongoing and improved performance is assured due to the vast number of applications that would benefit from the technology. SiPM design is also far more adaptable to the requirements of each application. Devices can be manufactured with tuned spectral response, different doping concentrations, optimised dark count rate/active area ratio, selected pixel and total area size, and lower power dissipation. Future devices will feature larger areas and a greater number of pixels with improved PDE and reduced cross-talk, and with higher rate capability. Produced using a standard metal-oxide-silicon (MOS) process, mass production is straightforward and fabrication costs are low.

## 5. Conclusion

Operation of the SensL 1000 series SiPM has been confirmed in liquid nitrogen. Following multiple thermal cycles the device consistently recovered its initial room temperature performance on warming and no mechanical trauma was observed at the conclusion of testing.

The device has predictable consistent gain from 25°C to -196°C for constant over-voltages between 1V and 4V. A gain of $3.6\times10^6$ at an over-voltage of 6.8V was recorded at -196°C, the maximum value not being reached. The breakdown voltage was found to reduce linearly from 28.2V at 25°C to 24.2V at -196°C, the coefficient measured as 18.3±2.5mV/°C.

The dark count rate has been shown to reduce exponentially with temperature at all bias voltages, a single photoelectron rate of 1MHz at 25°C for an over-voltage of 4V reducing to 40Hz at -196°C. Cross-talk probability, measured at 7% for an over-voltage of 4V, was found to be independent of temperature although linearly proportional to gain. Since the dark count rate is a function of gain, optimum working bias voltage should be selected so as to minimize noise contribution whilst maintaining a clearly discernible signal.

The photon detection efficiency was assessed as a function of photon wavelength and temperature. For an over-voltage of 4V, the PDE, found again to be invariant with temperature, was measured as 25% for 460nm photons and 11% for 680nm photons. In all fundamental aspects, the performance linearity of the device was unaffected by temperature, onset of saturation found to occur at identical photon fluxes irrespective of the device temperature.

Results obtained indicate that the low temperature gain of the device is well suited for light detection in cryogenic environments, its photon detection efficiency is well matched to liquid argon targets featuring the waveshifter tetraphenyl butadiene (TPB), and in general is a good match for medium and high energy physics applications such as neutrino detection, large arrays of SiPMs offering a promising approach to optical readout of cryogenic liquid noble gas systems.

**Acknowledgements**

The authors would like to thank Mr Mark Ward for his valuable advice during the data collection phase.

**References**

[1] J. Angle et al., First Results from the XENON10 Dark Matter Experiment at the Gran Sasso National Laboratory, 2008 Phys. Rev. Lett. **100**, 021303

[2] G.J. Alner et al., First limits on WIMP nuclear recoil signals in ZEPLIN-II: A two-phase xenon detector for dark matter detection, 2007 Astropart. Phys. **28**, Issue 3, 287-302

[3] S. Amerio et al., Design, construction and tests of the ICARUS T600 detector, 2004 Nucl. Instr. and Meth., **A527**, 329-410

[4] M. J. Carson, J. C. Davies, E. Daw, R. J. Hollingworth, V. A. Kudryavtsev, T. B. Lawson, P. K. Lightfoot, J. E. McMillan, B. Morgan, S. M. Paling, M. Robinson, N. J. C. Spooner, D. R. Tovey, Neutron background in large-scale xenon detectors for dark matter searches, 2004 Astropart. Phys. **21**, 667-687

[5] P. Buzhan, B. Dolgoshein, L. Filatov, A. Ilyin, V. Kantzerov, V. Kaplin, A. Karakash, F. Kayumov, S. Klemin, E. Popova, S. Smirnov, Silicon photomultiplier and its possible applications, 2003 Nucl. Instr. and Meth. **A504**, 48-52

[6] V. Andreev et al., A high-granularity scintillator calorimeter readout with silicon photomultipliers, 2005 Nucl. Instr. and Meth. **A540**, 368-380

[7] L.C.C. Coelho, J.A.M. Lopes, D.S. Covita, A.S. Conceição, J.M.F. dos Santos, Xenon GPSC high-pressure operation with large-area avalanche photodiode readout, 2007 Nucl. Instr. and Meth. **A575**, 444-448

[8] L. Ludhova et al., Planar LAAPDs: temperature dependence, performance, and application in low-energy X-ray spectroscopy, 2005 Nucl. Instr. and Meth. **A540**, 169-179

[9] M. Moszynski, M. Szawlowski, M. Kapusta, M. Balcerzyk, Avalanche photodiodes in scintillation detection, 2003 Nucl. Instr. and Meth. **A497**, 226-233

[10] K. Ni, E. Aprile, D. Day, K.L. Giboni, J.A.M. Lopes, P. Majewski, M. Yamashita, Performance of a large area avalanche photodiode in a liquid xenon ionization and scintillation chamber, 2005 Nucl. Instr. and Meth. **A551**, 356-363

[11] G. Bondarenko, P. Buzhan, B. Dolgoshein, V. Golovin, E. Guschin, A. Ilyin, V. Kaplin, A. Karakash, R. Klanner, V. Pokachalov, E. Popova, K. Smirnov, Limited Geiger-mode microcell silicon photodiode: new results, 2000 Nucl. Instr. and Meth. **A442**, 187-192

[12] E. Aprile, P. Cushman, K. Ni, P. Shagin, Detection of liquid xenon scintillation light with a silicon photomultiplier, 2006 Nucl. Instr. and Meth., **A556**, 215-218

[13] E. Lorenz, Status of the 17 m MAGIC telescope, 2004 New Astron. Rev.**48**, 339-344

[14] P. Buzhan, B. Dolgoshein, E. Garutti, M. Groll, A. Karakash, V. Kaplin, V. Kantserov, F. Kayumov, S. Klyomin, N. Kondratieva, A. Pleshko, E. Popova, F. Sefkow, Timing by silicon photomultiplier: A possible application for TOF measurements, 2006 Nucl. Instr. and Meth. **A567**, 353-355

[15] Qingguo Xie; R.G. Wagner, G. Drake, P. De Lurgio, Yun Dong, Chin-Tu Chen, Chien-Min Kao, Performance evaluation of multi-pixel photon counters for PET imaging, 2007 Nuclear Science Symposium Conference Record, IEEE, Vol. **2**, 969-974

[16] A.N. Otte, J. Barral, B. Dolgoshein, J. Hose, S. Klemin, E. Lorenz, R. Mirzoyan, E. Popova, M. Teshima, A test of silicon photomultipliers as readout for PET, 2005 Nucl. Instr. And Meth. **A545**, 705-715

[17] W.G. Oldham, R.R. Samuelson, P. Antognetti, Triggering phenomena in avalanche photodiodes, 1972 IEEE TED, Vol. **19**, Issue 9, 1056-1060

[18] Mineev et al., Scintillator counters with multi-pixel avalanche photodiode readout for the ND280 detector of the T2K experiment, 2007 Nucl. Instr. and Meth. **A577**, 540-551

[19] K. Mavrokoridis, P.K. Lightfoot, N.J.C. Spooner, Development and optimisation of the wavelength shifter coating and reflector combination for the the ArDM argon dark matter detector, in preparation.

[20] J.S. Kapustinsky, R.M. DeVries, N.J. DiGiacomo, W.E. Sondheim, J.W. Sunier, H. Coombes, A fast timing light pulser for scintillation detectors, 1985 Nucl. Instr. and Meth. **A241**, 612-613

[21] S. Uozumi, Study and Development of Multi Pixel Photon Counter for the GLD Calorimeter Readout, International workshop on new photon-detectors PD07, Kobe University, Kobe, Japan, 27-29 June, 2007  http://pos.sissa.it//archive/conferences/051/022/PD07_022.pdf

[22] B. Dolgoshein, et al., Status report on silicon photomultiplier development and its applications, 2006 Nucl. Instr. and Meth. **A563**, 368-376

[23] A. Lacaita, M. Ghioni, F. Zappa, G. Ripamonti, S. Cova, Recent advances in the detection of optical photons with silicon photodiodes, 1993 Nucl. Instr. and Meth. **A326**, 290-294

[24] K. Deiters, Y. Musienko, S. Nicol, B. Patel, D. Renker, S. Reucroft, R. Rusack, T. Sakhelashvili, J. Swain, P. Vikas, Properties of the most recent avalanche photodiodes for the CMS electromagnetic calorimeter, 2000 Nucl. Instr. and meth. **A442**, 193-197

[25] R. Lecomte, C. Pepin, D. Rouleau, H. Dautet, R. J. McIntyre, D. McSween, P. Webb, Radiation detection measurements with a new "Buried Junction" silicon avalanche photodiode, 1999 Nucl. Instr. and Meth. **A423**, 92-102

[26] A. Afanasjev et al., Scintillator counters with multi-pixel avalanche photodiode readout for the ND280 detector of the T2K experiment, 2007 Nucl. Instr. and Meth. **A577**, 540-551

[27] J. Ninković, Recent developments in silicon photomultipliers, 2007 Nucl. Instr. and Meth. **A580**, 1020-1022

[28] I. Britvitch, K. Deiters, Q. Ingram, A. Kuznetsov, Y. Musienko, D. Renker, S. Reucroft, T. Sakhelashvili, J. Swain, Avalanche photodiodes now and possible developments, 2004 Nucl. Instr. and Meth. **A535**, 523-527

[29] K. Yamamoto, K. Yamamura, K. Sato, S. Kamakura, T. Ota, H. Suzuki, S. Ohsuka, Newly developed semiconductor detectors by Hamamatsu, International workshop on new photon-detectors PD07, Kobe University, Kobe, Japan, 27-29 June, 2007 http://pos.sissa.it/archive/conferences/051/004/PD07_004.pdf